\documentclass[twocolumn,showpacs,preprintnumbers,amsmath,amssymb]{revtex4}

\usepackage{graphicx}
\usepackage{dcolumn}
\usepackage{bm}
\usepackage{tikz}
\usetikzlibrary{arrows,automata}
\usepackage{graphicx}
\usepackage{dcolumn}

\usepackage{amssymb}
\usepackage{amsmath}

\usepackage{booktabs}
\usepackage{mathtools}
\usepackage{array}

\begin{document}

\preprint{APS/123-QED}

\title{Steady states of continuous-time open quantum walks}

\author{Chaobin Liu}
\email{cbliu2000@yahoo.com}
\affiliation{Department of Mathematics, Bowie State University, Bowie, Maryland 20715, USA}
\author{Radhakrishnan Balu}
\email{radhakrishnan.balu.civ@mail.mil}
\affiliation{U.S. Army Research Laboratory, Computational and Information Sciences Directorate, Adelphi, Maryland 20783, USA}


\begin{abstract}

Continuous-time open quantum walks (CTOQW) are introduced as the formulation of quantum dynamical semigroups of trace-preserving and completely positive linear maps (or quantum Markov semigroups) on graphs. We show that a CTOQW always converges to a steady state regardless of the initial state when a graph is connected. When the graph is both connected and regular, it is shown that the steady state is the maximally mixed state. As shown by the examples in this article, the steady states of CTOQW can be very unusual and complicated even though the underlying graphs are simple. The examples demonstrate that the structure of a graph can affect quantum coherence in CTOQW through a long time run. Precisely, the quantum coherence persists throughout the evolution of the CTOQW when the underlying topology is certain irregular graphs (such as a path or a star as shown in the examples). In contrast, the quantum coherence will eventually vanish from the open quantum system when the underlying topology is a regular graph (such as a cycle). 



\end{abstract}

\maketitle

\section{Introduction}
A standard starting point for the discussion of the dynamics of continuous-time open quantum systems is the Markovian quantum master equation \cite{GKS1976, L1976, BP2006}

\begin{eqnarray}
\!\!\!\frac{d\rho(t)}{dt}
\!\!&=&\!\!i[\rho(t),H(t)]+\sum_k\gamma_k(t)[L_k(t)\rho(t)L^{\dagger}_k(t)\nonumber \\
\!\!&-&\!\!\frac{1}{2}\{L^{\dagger}_k(t)L_k(t),\rho(t)\}].
\end{eqnarray}
Here, $\rho(t)$ is the state of the system at time $t$, the self-adjoint operator $H(t)$ is the generator of the coherent part of the evolution, and the Lindblad operators $L_k(t)$ are the incoherent or dissipative system operators of the evolution with corresponding relaxation times $\gamma_k(t)$.

When $H(t)$, $L_k(t)$ and $\gamma_k(t)$ are independent of time $t$, the Markovian master equation describes a homogeneous evolution of continuous-time open quantum system. For simplicity, all $\gamma_k(t)$ can be assumed to be 1, then the above equation is written as
\begin{eqnarray}
\!\!\!\frac{d\rho(t)}{dt}
\!\!&=&\!\!i[\rho(t),H]+\sum_k[L_k\rho(t)L^{\dagger}_k\nonumber \\
\!\!&-&\!\!\frac{1}{2}\{L^{\dagger}_kL_k,\rho(t)\}]. \label{master}
\end{eqnarray}

Introducing an operator $\mathcal{L}$ as follows
\begin{eqnarray}
\mathcal{L}\rho=i[\rho,H]+\sum_k[L_k\rho L^{\dagger}_k-\frac{1}{2}\{L^{\dagger}_kL_k,\rho\}], \label{generator_L}
\end{eqnarray}
one can rewrite Eq. (\ref{master}) as 

\begin{eqnarray}
\frac{d}{dt}\rho(t)=\mathcal{L}\rho(t).\label{master-1}
\end{eqnarray}

As we may know, $T_t=e^{t\mathcal{L}}$ forms a quantum dynamical semigroup of trace-preserving and completely positive linear maps, which is continuous in $t\in \mathbb{R}_{+}$. For any density matrix $\rho(0)$, $\rho(t)=T_t(\rho(0))$ fulfills Eq.(\ref{master-1}). A detailed justification for these assertions can be found in literature, for example \cite{Wolf2010}. In the context of quantum dynamical semigroup the operator $\mathcal{L}$ is often called the generator or infinitesimal generator of the semigroup.

When dynamical processes take place on a graph, the topology of the graph often affects or even determines the behavior of the processes \cite{AF2002, MB2011}. Our work here is concerned with a semigroup taking place on a graph; we are interested in determining the limiting states of the semigroup and knowing how the structure of the graph can determine the behavior of the semigroup. The study shows through the examples of this article, among other things, that the long time behavior of the semigroups can be very unusual and complicated even though the underlying graphs are simple. The aforesaid key attribute of the semigroups on graphs may be useful for quantum information processing and quantum computing \cite{NC2000}. 



To be more precise, we now consider a scenario that the quantum dynamical semigroup governed by Eq.(\ref{master-1}) is confined to a graph $G=(V, E)$. Here, $\rho(t)$, the state of the system at time $t$, is a density matrix on the Hilbert space spanned by the orthonormal basis represented by the vertices of $G$. We interpret $H$, the generator of the coherent part, to be the {\it Laplacian} matrix of $G$, and $L_k$, the dissipative system operators, to be operators determined by the adjacency matrix of $G$ (the precise description of these operators will be given in section II). This interpretation is not necessary. Nonetheless, it respects the topological structure of a graph $G$.

In this article, the formulation of the quantum dynamical semigroup of trace-preserving, completely positive linear maps on a graph $G$ will be called {\it continuous-time open quantum walks} (abbreviated to CTOQW)\cite{WRA10, LB2017}.

We remark that a different approach to the notion of the CTOQW was recently proposed \cite{Pell2014}. This type of continuous-time open quantum walks is obtained by taking a continuous time limit of the discrete-time open quantum random walks \cite{APS12, APSS12}. More recently, a second different species of CTOQW was presented \cite{SP2015}, it can be considered as the exact continuous-time version of the aforesaid discrete-time open quantum walks 


It is noteworthy that there are two extreme cases of possible continuous time transport processes on graphs that have been extensively studied. One is continuous time random walks (CTRW) modeling a purely incoherent transport process, see, e.g., \cite{MB2011}. The other is continuous time quantum walks (CTQW) modeling a purely coherent transport process \cite{FG1998}. The probability distribution for CTRW at time $t$ is given by $p(t)=e^{tL}p(0)$ where $p(0)$ is the initial probability distribution. The transition amplitude for CTQW at time $t$ is given by $|\psi\rangle=e^{-itL}|\psi(0)\rangle$ where $|\psi(0)\rangle$ is the initial amplitude of CTQW. In both cases described above, $L$ can be chosen as the {\it Laplacian} matrix of the graph. The history and developments of continuous-time quantum walks, and their application to various sciences can be found in \cite{K06, MB2011, Salvador2012} and the references therein.

Coming back to CTOQW to be formulated, we may consider it as a mixture of the two aforesaid extreme cases of continuous-time transport processes on graphs to some extent. In addition to formulating CTOQW, we will also study properties of steady states of CTOQW. 

This article is organized as follows. In section II, the formalism for continuous-time open quantum walks is presented. We then analyze the steady states of CTOQW and offer examples of CTOQW in section III. This article is closed in section IV with concluding remarks and related questions.


\section{Formulation of continuous-time open quantum walks}

\subsection{Matrices associated to a graph}

In this section, we will introduce continuous-time open quantum walks (CTOQW) as the formulation of homogeneous quantum Markov semigroups on graphs.


To prepare for the formulation, we will first recall some matrices associated to a graph. Given an undirected graph $G=(V, E)$ without multiple edges or self-loops, let $A$ be the {\it adjacency matrix} of G, the $|V|\times |V|$ matrix with the elements

$A_{jk}=\left\{
\begin{array}{l}
1$ if $(j,k)\in E\\
0$ if $(j,k)\notin E
\end{array}\right. $
for every pair $j, k\in V$. 


We denote by $D$ the degree matrix with $D_{jj}=\mathrm{deg}(j)$. Here, $\mathrm{deg}(j)$ denotes the degree of vertex $j$ for each vertex $j$ in $V$.
There are two more equally important matrices associated to the graph $G$: the {\it Laplacian} matrix of G, and the (canonical) {\it matrix of transition probabilities} for the Markov chain generated by the graph $G$.

Their elements are defined as:

\vskip 0.1in
$L_{jk}=\left\{
\begin{array}{l}
$deg$(j)$ if $j=k\\
-1$ if $(j,k)\in E\\
0$ if $(j,k)\notin E
\end{array}\right. $

\vskip 0.1in
$M_{jk}=\left\{
\begin{array}{l}
\frac{1}{\mathrm{deg}(k)}$ if $(j,k)\in E\\
0$ Otherwise $
\end{array}\right. $

for every pair $j, k\in V$. By their definitions, it is evident that $L=D-A$, and $M=AD^{-1}$.

One can associate with every vertex of the graph G a basis vector in an $|V|$-dimensional vector space. Now, these basis vectors form a complete orthonormal basis, which, for instance, is given by

\begin{equation}
 |1\rangle =\left[\begin{array}{c}
  1\\
   0\\
    \vdots\\
     0
     \end{array}\right],
 |2\rangle =\left[\begin{array}{c}
  0\\
   1\\
    \vdots\\
     0
     \end{array}\right],\ldots,
 |N\rangle =\left[\begin{array}{c}
  0\\
   0\\
    \vdots\\
     1
     \end{array}\right].
\end{equation}

Take as an example a cycle of 3 sites, then the matrices $L$ and $M$ read:
\begin{equation}
L= \left[\begin{array}{ccc}
2& -1 &-1 \\
-1      & 2 & -1\\
-1 & -1& 2
\end{array}\right],
M= \left[\begin{array}{ccc}
0& 1/2 &1/2 \\
1/2      & 0 & 1/2\\
1/2 & 1/2& 0
\end{array}\right].
\end{equation}

Writing $M$ in a quantum mechanical fashion via the projection operators $|j\rangle \langle k|$, leads to 

\begin{eqnarray}
\!\!\!M
\!\!&=&\!\!\frac{1}{2}\sum_{j\ne k}|k\rangle\langle j|\nonumber\\
\!\!&=&\!\!\frac{1}{2}[|2\rangle\langle 1|+|3\rangle\langle 1|+|1\rangle\langle 2|+|3\rangle\langle 2|+|1\rangle\langle 3|+|2\rangle\langle 3|]
\end{eqnarray}

\subsection{Definition of continuous-time open quantum walks}

Having now specified a graph $G$ through its associated matrices, we are in a position to define the generator of the semigroup $e^{t\mathcal{L}}$ on $G$. We choose $H$ in Eq.(\ref{generator_L}) to be the {\it Laplacian} matrix of the graph $G$, Lindblad operators to be the swap operator $B_{jk}=\sqrt{\mathrm{M}_{jk}}|j\rangle\langle k|$. The parameters $\mathrm{M}_{jk}$ specify the strength of swapping the state $|k\rangle$ for the state $|j\rangle$.

Then the generator $\mathcal{L}$ can be recast as

\begin{eqnarray}
\mathcal{L}\rho=i[\rho,L]+\sum_{j,k}[B_{jk}\rho B_{jk}^{\dagger}-\frac{1}{2}\{B_{jk}^{\dagger}B_{jk},\rho\}].\label{generatorCTOQW}
\end{eqnarray}

For a given initial state $\rho(0)$ of the quantum system on $G$, the expression $\rho(t)=T_t(\rho(0))=e^{t\mathcal{L}}\rho(0)$ is called the state of the continuous-time open quantum walk on $G$ at time $t$. 

It should be pointed out that the group of Lindblad operators given in Eq. (\ref{generatorCTOQW}) is just one of possible options; there are other options for Lindblad operators in Eq. (\ref{generator_L}). For example, 1) the Lindblad operators can be chosen to be $e^{-iL}$ and $\mathbb{I}$ where $L$ is the {\it Laplacian} matrix of a graph; or 2) they can be chosen to be a projection operator $P$ and $\mathbb{I}-P$ where $P=|1\rangle\langle 1|+...+...+|j\rangle \langle j|+...+|m\rangle\langle m|$ with $j$ standing for the $j$th vertex of a graph. In what follows, unless specified, the Lindblad operators to be used are ones given by Eq.(\ref{generatorCTOQW}).


\section{Steady states of continuous-times open quantum walks}

In this section, we will study properties of steady states of CTOQW. In the theory of completely positive semigroups, a semigroup $T_t$ is said to be relaxing if there exists a unique (steady) state $\rho_{\infty}$ such that $T_t(\rho_{\infty})=\rho_{\infty}$ for all $t$ and
$\lim_{t\rightarrow \infty} T_t(\rho)=\rho_{\infty}$ for an arbitrary state $\rho$.

It can be verified that $\rho$ is a steady state of $T_t$ for $t>0$ if and only if $\rho$ is in the kernel of $\mathcal{L}$, i.e., $T_t(\rho)=\rho$ for all $t>0$ if and only if $\mathcal{L}(\rho)=0$. The semigroup is obviously not relaxing if the equation $\mathcal{L}(\rho)=0$ admits more than one solution.

It is desirable to have some condition on the structure of the graph which assures that the semigroup defining CTOQW is relaxing as well
as a characterization of the steady states.

Here, we recall a very important result (please refer to \cite{RH12, Wolf2010} and the references cited therein) as a sufficient condition for a semigroup being relaxing.




\noindent Lemma 1. Consider a quantum dynamical semigroup of trace-preserving, completely positive linear maps, $T_t=e^{\mathcal{L}t}$ with generator
\begin{eqnarray}
\mathcal{L}\rho=i[\rho,H]+\sum_k[L_k\rho L^{\dagger}_k-\frac{1}{2}\{L^{\dagger}_kL_k,\rho\}], 
\end{eqnarray}
for some set of indices $I$. Provided that the linear space spanned by the set of Lindblad operators $L_k$ is Hermitian (this means that for every $X =\sum_kx_kL_k$ there are $y_k\in \mathbb{C}$ such that $X^{\dagger}=\sum_k y_kL_k$.) and the only operators commuting with all of them are proportional to the identity, which is expressed as $\{L_k, k\in I\}^{\prime}=c\mathbb{I}$ (Note: the left-hand side of this equality is called the commutant of $\{L_k, k\in I\}$), then the semigroup $T_t$ is relaxing and the steady state is a positive definite density matrix.

We now present our basic result which gives a sufficient condition for convergence of a continuous-time open quantum walk. 


\noindent Theorem 1.\, If a graph $G$ is connected, then CTOQW on the graph has a positive definite matrix (state) $\rho_{\infty}$ such that for all density matrices $\rho$ we have $\lim_{t\rightarrow\infty}e^{t\mathcal{L}}\rho=\rho_{\infty}$. If $G$ is connected, then $\lim_{t\rightarrow\infty}e^{t\mathcal{L}}\rho=\frac{1}{|V|}\mathbb{I}$ for any initial state $\rho$ if and only if the associated matrix of transition probabilities on $G$ is doubly stochastic.

We call a square matrix with nonnegative entries doubly stochastic if the sum of the entries in each row and in each column is 1.

\noindent Proof.\, Since G is an undirected graph, $\mathrm{M}_{jk}\neq 0$ iff $\mathrm{M}_{kj}\neq 0$ for any $j$ and $k$. This implies that the linear space spanned by the set of Lindblad operators $B_{jk}$ is Hermitian. Next we prove that $\{B_{jk}, j,k\in I\}^{\prime}=c\mathbb{I}$. Provided that $X=(X_{lm})\in \{B_{jk}, j,k\in I\}^{\prime}$, i.e.,  $B_{jk}X=XB_{jk}$ for all possible $B_{jk}$, then we claim $X=c\mathbb{I}$ for some constant $c$. To justify it, we will show that for each $l_0$,  $X_{l_0m_0}=0$ whenever $m_0\ne l_0$. Now for the given $l_0$, there exists $k_0$ such that $B_{k_0l_0}\neq 0$ as G is connected. Please note that 

\begin{eqnarray}
\!\!\!B_{k_0l_0}X
\!\!&=&\!\!\frac{1}{\sqrt{\mathrm{deg}(l_0)}}|k_0\rangle\langle l_0|\sum_{j,m}X_{jm}|j\rangle\langle m|\nonumber\\
\!\!&=&\!\!\frac{1}{\sqrt{\mathrm{deg}(l_0)}}\sum_{m}X_{l_0m}|k_0\rangle\langle m|,\nonumber\\
\!\!\!XB_{k_0l_0}
\!\!&=&\!\!\frac{1}{\sqrt{\mathrm{deg}(l_0)}}\sum_{j,m}X_{jm}|j\rangle\langle m|k_0\rangle\langle l_0|\nonumber\\
\!\!&=&\!\!\frac{1}{\sqrt{\mathrm{deg}(l_0)}}\sum_{j}X_{jk_0}|j\rangle\langle l_0|.
\end{eqnarray}

Then $B_{k_0l_0}X=XB_{k_0l_0}$ implies that $X_{l_0m_0}=0$ whenever $m_0\ne l_0$ and $X_{l_0l_0}=X_{k_0k_0}$. Repeated applying this reasoning leads to the claim that $X=c\mathbb{I}$ where $c$ is a constant. Thus $\{B_{jk}, j, k\in I\}^{\prime}=c\mathbb{I}$. By Lemma 1, CTOQW on the graph has a positive definite matrix ( steady state ) $\rho_{\infty}$ such that for all density matrices $\rho$ we have $\lim_{t\rightarrow\infty}e^{t\mathcal{L}}\rho=\rho_{\infty}$.

To prove the second assertion, it suffices to show that $\mathcal{L}(\mathbb{I})=0$ if and only if $\sum_kM_{jk}=1$ for all $j$ (which means that $M$ is doubly stochastic).

Since the sum of all entries on each column of $M$ is one, i.e. $\sum_{j}M_{jk}=1$ for all $k$, we have 
\begin{eqnarray}
\!\!\!\sum_j\sum_k B_{jk}^{\dagger}B_{jk}
\!\!&=&\!\!\sum_k\sum_j(\sqrt{M_{jk}})^2|k\rangle\langle j|j\rangle\langle k|\nonumber \\
\!\!&=&\!\!\sum_k(\sum_jM_{jk})|k\rangle\langle k|\nonumber\\
\!\!&=&\!\!\mathbb{I}. \label{b_1}
\end{eqnarray}

Observe that
\begin{eqnarray}
\!\!\!\sum_j\sum_k B_{jk}B_{jk}^{\dagger}
\!\!&=&\!\!\sum_j\sum_k(\sqrt{M_{jk}})^2|j\rangle\langle k|k\rangle\langle j|\nonumber \\
\!\!&=&\!\!\sum_j(\sum_kM_{jk})|j\rangle\langle j|.\label{b_2}
\end{eqnarray}

Eqs. (\ref{b_1}) and (\ref{b_2}) together imply the following identities:
\begin{eqnarray}
\!\!\!\mathcal{L}(\mathbb{I})
\!\!&=&\!\! i[\mathbb{I},L]+\sum_{j,k}[B_{jk}\mathbb{I} B_{jk}^{\dagger}-\frac{1}{2}\{B_{jk}^{\dagger}B_{jk},\mathbb{I}\}]\nonumber\\
\!\!&=&\!\!\sum_j(\sum_kM_{jk})|j\rangle\langle j|-\mathbb{I}.
\end{eqnarray}

Note that $\mathcal{L}(\mathbb{I})=0$ iff $\sum_j(\sum_kM_{jk})|j\rangle\langle j|=\mathbb{I}$ iff $\sum_kM_{jk}=1$ for all $j$.

Therefore the result follows.

\vskip 0.2in

Since the matrix of transition probabilities associated with a regular graph is doubly stochastic, Theorem 1 has an immediate corollary, which concerns the convergence of CTOQW on a regular graph.

\vskip0.2in

\noindent Corollary 1.\, If graph $G$ is regular, then  $\lim_{t\rightarrow\infty}e^{t\mathcal{L}}\rho=\frac{1}{|V|}\mathbb{I}$ for any initial state $\rho$.

\vskip0.2in

In what follows, we consider three examples of CTOQW. The first example of CTOQW is the one on an $n-$cycle. Since the graph is regular, $\frac{1}{n}\mathbb{I}$ is the unique steady state such that $\lim_{t\rightarrow\infty}e^{t\mathcal{L}}\rho=\frac{1}{n}\mathbb{I}$ for any initial state $\rho$ according to Corollary 1.
 
We then consider the CTOQW on a path with three sites (see the graph below). Meanwhile, we also consider the corresponding CTRW and CTQW on this graph to compare the long time behaviors of these three distinct continuous-time dynamical processes in terms of steady states and limiting probability distributions.



\begin{figure}[h!]
  \begin{center}
   \begin{tikzpicture}
\tikzstyle{every node}=[draw,shape=circle];
\node (v1) at (0,0) {$v_1$};
\node (v2) at ( 2,0) {$v_2$};
\node (v3) at ( 4,0) {$v_3$};
\draw (v1) -- (v2)
(v2) -- (v3);
\end{tikzpicture} 
    \caption{A path with three sites}
  \end{center}
\end{figure}
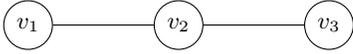

The Laplacian matrix and the matrix of transition probabilities associated with this graph are as follows:
\begin{equation}
L=\left[\begin{array}{ccc}
1& -1& 0\\
-1&2&-1\\
0& -1&1
\end{array}\right],\, M=\left[\begin{array}{ccc}
0& 1/2& 0\\
1&0&1\\
0& 1/2&0
\end{array}\right] 
\end{equation}

The Lindblad operators/Swap operators are given by:

\begin{equation}
B_{12}=\sqrt{2}/2|v_1\rangle\langle v_2|=\left[\begin{array}{ccc}
0& \sqrt{2}/2& 0\\
0&0&0\\
0& 0&0
\end{array}\right]
\end{equation}
\begin{equation}
B_{32}=\sqrt{2}/2|v_3\rangle\langle v_2|=\left[\begin{array}{ccc}
0& 0& 0\\
0&0&0\\
0& \sqrt{2}/2&0
\end{array}\right]
\end{equation}

\begin{equation}
B_{21}=|v_2\rangle\langle v_1|=\left[\begin{array}{ccc}
0& 0& 0\\
1&0&0\\
0& 0&0
\end{array}\right]
\end{equation}

\begin{equation}
B_{23}=|v_2\rangle\langle v_3|=\left[\begin{array}{ccc}
0& 0& 0\\
0&0&1\\
0& 0&0
\end{array}\right]
\end{equation}

\vskip.2in

Since this path is a connected graph, CTOQW has a unique steady state by Theorem 1. To find this steady state, it suffices to solve for $\rho$ the system of linear equations of $\mathcal{L}(\rho)=0$ and $\sum^3_{j=1}\rho_{jj}=1$. Here, $\mathcal{L}$ is given by Eq. (\ref{generatorCTOQW}) and $\rho=(\rho_{jk})_{3\times 3}$. After massive calculation, we obtain a unique solution to the system of the linear equations, which is given by

\begin{equation}
\left[\begin{array}{ccc}
\frac{2}{7}& -\frac{1}{28}+\frac{1}{28}i& \frac{1}{14}\\
-\frac{1}{28}-\frac{1}{28}i&\frac{3}{7}&-\frac{1}{28}-\frac{1}{28}i\\
\frac{1}{14}& -\frac{1}{28}+\frac{1}{28}i&\frac{2}{7}
\end{array}\right] \label{stationarystate}
\end{equation}

It can be verified that the above matrix is a positive definite density matrix. To distinguish this steady state from other density states, we denote it by $\rho_{\infty}$. According to Theorem 1, for CTOQW on this path, we have $\lim_{t\rightarrow\infty}e^{t\mathcal{L}}\rho=\rho_{\infty}$ for any initial state $\rho$. This implies that the limiting probability distribution over the three sites, namely $v_1$, $v_2$ and $v_3$, is $(\frac{2}{7},\frac{3}{7}, \frac{2}{7})$ regardless of the initial state of CTOQW.

\vskip 0.2in

We now turn to the two aforesaid extreme cases of continuous-time processes on this graph. For CTRW, it is interesting to note that the long-time limit of all transition probabilities is $\lim_{t\rightarrow \infty}p(t) = 1/3$. In fact, every CTRW whose transfer matrix follows directly from the {\it Laplacian} matrix will eventually decay at long times to the equipartition value $1/N$, and is independent of the connectivity of the graph \cite{MB2011}.

\vskip 0.2in

For CTQW, the eigenvalues and the corresponding unit eigenvectors of {\it Laplacian} matrix $L$ are as follows:

\begin{eqnarray}
\lambda_1\!\!&=&\!\!0, \psi_1=(\frac{\sqrt{3}}{3}, \frac{\sqrt{3}}{3}, \frac{\sqrt{3}}{3})^T\\
\lambda_2\!\!&=&\!\!1, \psi_2=(-\frac{\sqrt{2}}{2},0,\frac{\sqrt{2}}{2})^T\\
\lambda_3\!\!&=&\!\!3, \psi_3=(\frac{\sqrt{6}}{6},-\frac{\sqrt{6}}{3},\frac{\sqrt{6}}{6})^T
\end{eqnarray}

By standard computation rule, the limiting average probability of finding the quantum walker at a site $v$ initially launching at a site $u$ is given by $P_{\infty}(v|u)=\sum_{j=1}^3|\langle v|\psi_j\rangle|^2|\langle u|\psi_j\rangle|^2$.

We now show the limiting average probability distributions of the CTQW launching from two specific states:

Case 1: when the walker is launched at $v_1$ (or $v_3$), then the limiting average probability distribution is $(\frac{7}{18}, \frac{4}{18},\frac{7}{18})$.

Case 2: when the walker is launched at $v_2$, then the limiting average probability distribution is $(\frac{2}{9},\frac{5}{9},\frac{2}{9})$.

As shown above, the asymptotic behaviors of these three continuous-time processes on the path are quite distinct. Unlike CTQW, both CTOQW and CTRW have unique limiting distributions which are independent of initial states/probability distributions. 
\vskip 0.2in

Finally we consider the CTOQW on a star with three edges (this graph is often called a claw and is shown below).

\begin{figure}[h!]
  \begin{center}
   \begin{tikzpicture}
\tikzstyle{every node}=[draw,shape=circle];
\node (v1) at (0,0) {$v_1$};
\node (v2) at (0,-2) {$v_2$};
\node (v3) at (1.7 ,1) {$v_3$};
\node (v4) at (-1.7, 1) {$v_4$};
\draw (v1) -- (v2)
(v1) -- (v3)
(v1)--(v4);
\end{tikzpicture} 
    \caption{A star with 3 edges}
  \end{center}
\end{figure}
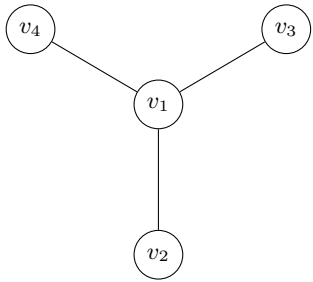

The Laplacian matrix and the matrix of transition probabilities associated with this graph are as follows:
\begin{equation}
L=\left[\begin{array}{cccc}
3& -1&-1&-1\\
-1&1&0&0\\
-1& 0&1&0\\
-1&0&0& 1
\end{array}\right],\, M=\left[\begin{array}{cccc}
0&1&1&1\\
1/3&0&0&0\\
1/3&0&0&0\\
1/3&0&0&0
\end{array}\right] 
\end{equation}

Since the star is connected, there is a unique steady state denoted by $\rho_{\infty}$ such that $\lim_{t\rightarrow\infty}e^{t\mathcal{L}}\rho=\rho_{\infty}$ for any initial state $\rho$ according to Theorem 1. By a similar approach employed in the second example, we obtain the steady state given by:

\begin{equation}
\rho_{\infty}=\left[\begin{array}{cccc}
\frac{11}{26}& -\frac{2}{39}-\frac{1}{39}i& -\frac{2}{39}-\frac{1}{39}i&-\frac{2}{39}-\frac{1}{39}i \\
-\frac{2}{39}+\frac{1}{39}i&\frac{5}{26}&\frac{2}{39}&\frac{2}{39}\\
-\frac{2}{39}+\frac{1}{39}i& \frac{2}{39}&\frac{5}{26}&\frac{2}{39}\\
-\frac{2}{39}+\frac{1}{39}i&\frac{2}{39}&\frac{2}{39}&\frac{5}{26}
\end{array}\right] \label{stationarystate_2}
\end{equation}

\vskip 0.2in


It should be noted that the limiting density matrices of CTOQW such as the ones given by Eqs.(\ref{stationarystate}) and (\ref{stationarystate_2}) can be very complicated and unusual even though the underlying graphs are simple. It also should be noted that the nonzeroness of the off-diagonal elements of the limiting density matrices indicates the existence of quantum coherence throughout the evolution of the CTOQW on the graphs. This finding suggests that quantum coherence persists throughout the evolution of the CTOQW when the underlying topology is certain irregular graphs (such as a path or a star as shown in the examples). In contrast, quantum coherence will eventually vanish from the open quantum system when the underlying topology is a regular graph (such as a cycle) by Theorem 1. 

\vskip 0.2in

\section{Concluding remarks and related questions}

In this article, we examine the evolution of CTOQW on finite graphs. In general, for dynamical processes on graphs,
one is interested not only in determining the shape of the limiting state (also known as steady state) of the processes and  the condition under which the limiting state exists,  but also in determining the relation between steady states and the structures of the underlying graphs. We have shown that a CTOQW always converges to a steady state regardless of the initial state when a graph is connected. When the graph is both connected and regular, it is shown that the steady state is the maximally mixed state. As shown by the examples in this article, the steady states of CTOQW can be very unusual and complicated even though the underlying graphs are simple. Our examples with the steady states of CTOQW, to our best knowledge, may be the first examples that were reported in literature for semigroups with explicit steady states determined by graphs.

CTOQW on graphs can be extended to lattices. For instance, consider CTOQW on one-dimensional lattice $\mathbb{Z}$. One may choose $H$ in Eq.(\ref{generator_L}) to be the {\it Laplacian} matrix of the lattice $\mathbb{Z}$, and choose the swap operator $B_{x,x-1}=\frac{\sqrt{2}}{2}|x\rangle\langle x-1|$ and $B_{x,x+1}=\frac{\sqrt{2}}{2}|x\rangle\langle x+1|$ as Lindblad operators. It would be interesting to investigate the asymptotic behavior of $\rho_t/\sqrt{t}$.

\vskip 0.3in

\section{ACKNOWLEDGMENTS}

We are pleased to acknowledge that we had useful communication with Carlos F. Lardizabal when we were working on this article. We thank undergraduate research assistant Carlos Sanchez for computing the steady state in the second example. CL gratefully acknowledges the support from ARL through ARL Faculty Fellow Research Team Program.

\end{document}